\documentclass[12pt]{article}
\pdfoutput=1

\usepackage{amssymb,amsmath,color,natbib,graphicx,
  setspace,sectsty,anysize,times,dsfont}

\usepackage[small]{caption}

\marginsize{1.1in}{.9in}{.3in}{1.4in}

\newcommand{\dbl}{\setstretch{1.6}}
\newcommand{\hlf}{\setstretch{1.2}}

\newcommand{\bs}[1]{\boldsymbol{#1}}
\newcommand{\mc}[1]{\mathcal{#1}}
\newcommand{\mr}[1]{\mathrm{#1}}
\newcommand{\bm}[1]{\mathbf{#1}}
\newcommand{\ds}[1]{\mathds{#1}}

\sectionfont{\noindent\normalfont\large\bf}
\subsectionfont{\noindent\normalfont\large\textit}

\begin{document}

\hlf

\pagestyle{empty}

\noindent {\Large \bf{Mixture  Modeling for Marked Poisson Processes}}
\vskip 2cm

\noindent{\large 
Matthew A. Taddy} \\
{\tt taddy@chicagobooth.edu}\\
{ \it The  University of Chicago Booth School of Business\\
5807 South Woodlawn Ave, Chicago, IL 60637, USA }

\vskip .5cm
\noindent{\large  Athanasios Kottas} \\
{\tt thanos@ams.ucsc.edu}\\
{ \it Department of Applied Mathematics and Statistics\\
University of California, Santa Cruz\\
1156 High Street, Santa Cruz, CA 95064, USA}

\vskip 2cm

{\small 
\noindent{\sc Abstract:} 
  We propose a general modeling framework for marked 
  Poisson processes observed over time or space. The modeling approach
  exploits the connection of the nonhomogeneous Poisson process
  intensity with a density function. Nonparametric Dirichlet process
  mixtures for this density, combined with nonparametric or
  semiparametric modeling for the mark distribution, yield flexible
  prior models for the marked Poisson process. In particular, we focus
  on fully nonparametric model formulations that build the mark
  density and intensity function from a joint nonparametric mixture,
  and provide guidelines for straightforward application of these
  techniques.  A key feature of such models is that they can yield
  flexible inference about the conditional distribution for
  multivariate marks without requiring specification of a complicated
  dependence scheme.  We address issues relating to choice of the
  Dirichlet process mixture kernels, and develop methods for prior
  specification and posterior simulation for full inference about
  functionals of the marked Poisson process. Moreover, we discuss a
  method for model checking that can be used to assess and compare
  goodness of fit of different model specifications under the proposed
  framework. The methodology is illustrated with simulated and real
  data sets. }

\vskip 2cm


\newpage
\dbl
\pagestyle{plain}

\section{Introduction} 

Marked point process data, occurring on either spatial or temporal
domains, is encountered in research for biology, ecology, economics, 
sociology, and numerous other disciplines. Whenever interest lies in the
intensity of event occurrences as well as the spatial or temporal
distribution of events, the data analysis problem will involve
inference for a non-homogeneous point process. Moreover, many
applications involve {\it marks} -- a set of random variables
associated with each point event -- such that the data generating
mechanism is characterized as a marked point process. In marketing,
for example, interest may lie in both the location and intensity of
purchasing behavior as well as consumer choices, and the data may be
modeled as a spatial point process with purchase events and product
choice marks. As another example, in forestry interest often lies in
estimating the wood-volume characteristics of a plot of land by
understanding the distribution and type of tree in a smaller subplot.
Hence, the forest can be modeled as a spatial point process with
tree events marked by trunk size and tree species.

Non-homogeneous Poisson processes (NHPPs) play a fundamental role in 
inference for data consisting of point event patterns 
\citep[e.g.,][]{
Gutt1995,MollWaag2004}, and marked NHPPs provide the natural model 
extension when the point events are accompanied by
random marks. One reason for the common usage of Poisson processes is
their general tractability and the simplicity of the associated data
likelihood. In particular, for a NHPP,
$\mr{PoP}(\mc{R},\lambda)$, defined on the observation window $\mc{R}$
with intensity $\lambda(\bm{x})$ for $\bm{x} \in \mc{R}$, which is a 
non-negative and locally integrable function for all bounded 
$\mc{B} \subseteq \mc{R}$, the following hold true:
\begin{itemize}
\item[$i$.] For any such $\mc{B}$, the number of points in $\mc{B}$,
$N(\mc{B}) \sim$ $\mr{Po}(\Lambda(\mc{B}))$, where $\Lambda(\mc{B})=$ 
$\int_\mc{B} \lambda(\bm{x}) d\bm{x}$ is the NHPP cumulative intensity
function.
\item[$ii$.] Given $N(\mc{B})$, the point locations within $\mc{B}$
  are i.i.d. with density $\lambda(\bm{x})/ \int_\mc{B}
    \lambda(\bm{x}) d\bm{x}$.
\end{itemize}
Here, $\mr{Po}(\mu)$ denotes the Poisson distribution with mean $\mu$.
Although $\mc{R}$ can be of arbitrary dimension, we concentrate on the
common settings of temporal NHPPs with $\mc{R} \subset \ds{R}^+$,
or spatial NHPPs where $\mc{R} \subset \ds{R}^2$.

This paper develops Bayesian nonparametric mixtures to model the
intensity function of NHPPs, and will provide a framework for combining 
this approach with flexible (nonparametric or semiparametric) modeling 
for the associated mark distribution. Since we propose fully nonparametric
mixture modeling for the point process intensity, but within the context
of Poisson distributions induced by the NHPP assumption, the nature of our 
modeling approach is {\it semiparametric}.
We are able to take advantage of the above formulation of the NHPP and 
specify the sampling density $f(\bm{x})=$ $\lambda(\bm{x})/\Lambda_{\mc{R}}$ 
through a Dirichlet process (DP) mixture model, where $\Lambda_{\mc{R}} \equiv$
$\Lambda(\mc{R})=$ $\int_\mc{R} \lambda(\bm{x}) d\bm{x}$ is the total 
integrated intensity.  
Crucially, items $i$ and $ii$ above imply that the likelihood for a NHPP
generated point pattern $\{\bm{x}_1,\ldots,\bm{x}_N\} \subset \mc{R}$ 
factorizes as
\begin{equation}\label{eqn:lhd}
\mr{p}\left(\{\bm{x}_i\}_{i=1}^N ; \lambda(\cdot) \right) \equiv
\mr{p}\left(\{\bm{x}_i\}_{i=1}^N ; \Lambda_{\mc{R}}, f(\cdot) \right) \propto
\Lambda_{\mc{R}}^{N} \exp(-\Lambda_{\mc{R}}) \prod_{i=1}^{N} f(\bm{x}_{i}),
\end{equation}
such that the NHPP density, $f(\cdot)$, and integrated intensity,
$\Lambda_{\mc{R}}$, can be modeled separately.  In particular, the DP
mixture modeling framework for $f(\cdot)$ allows for data-driven
inference about non-standard intensity shapes and quantification of
the associated uncertainty.

This approach was originally developed by \citet{KottSans2007} in the
context of spatial NHPPs with emphasis on extreme value analysis problems,
and has also been applied to analysis of immunological studies
\citep{JiMerlKeplWest2009} and neuronal data analysis 
\citep{KottasBehseta2010}.
Here, we generalize the mixture model to alternative kernel choices that 
provide for conditionally conjugate models and, in the context of temporal 
NHPPs, for monotonicity restrictions on the intensity function. However, 
in addition to providing a more general approach for intensity estimation, 
the main feature of this paper is an extension of the intensity mixture 
framework to modeling marked Poisson processes. Indeed, the advantage of 
a Bayesian nonparametric model-based approach will be most clear when it is 
combined with  modeling for the conditional mark distribution, thus providing 
unified inference for point pattern data.

General theoretical background on Poisson processes can be found, for instance, 
in \citet{Cres1993}, \citet{King1993}, and \citet{DaleVere2003}. \citet{Digg2003} 
reviews likelihood and classical nonparametric inference for spatial NHPPs, and 
\citet{MollWaag2004} discusses work on simulation-based inference for spatial 
point processes. 

A standard approach to (approximate) Bayesian inference for NHPPs is
based upon log-Gaussian Cox process models, wherein the random
intensity function is modeled on logarithmic scale as a Gaussian
process \citep[e.g.,][]
{MollSyveWaag1998,BrixDigg2001,BrixMoll2001}. In particular,
\citet{LianCarlGelf2009} present a Bayesian hierarchical model for
marked Poisson processes through an extension of the log-Gaussian Cox
process to accommodate different types of covariate information.
Early Bayesian nonparametric modeling focused on the cumulative
intensity function, $\int_{0}^{t} \lambda(s) ds$, for temporal point
processes, including models based on gamma, beta or general L\'evy
process priors
\citep[e.g.,][]{Hjor1990,Lo1992,KuoGhos1997,GutiNiet2003}.  An
alternative approach is found in \citet{HeikArja1998,HeikArja1999},
where piecewise constant functions, driven by Voronoi tessellations
and Markov random field priors, are used to model spatial NHPP
intensities.

The framework considered herein is more closely related to approaches
that involve a mixture model for $\lambda(\cdot)$. In particular,
\citet{LoWeng1989} and \citet{IshwJame2004} utilize a mixture
representation for the intensity function based upon a convolution of
non-negative kernels with a weighted gamma process. Moreover,
\citet{WolpIcks1998} include the gamma process as a special case of
convolutions with a general L\'evy random field, while
\citet{IcksWolp1999} and \citet{BestIcksWolp2000} describe extensions
of the gamma process convolution model to regression settings.
\citet{IcksWolp1999} also provide a connection to modeling for marked
processes through an additive intensity formulation.  
Since these mixture models have the integrated intensity term linked 
to their nonparametric prior for $\lambda(\cdot)$, they can be cast 
as a generalization of our model of independent $\Lambda_{\mc{R}}$.

A distinguishing feature of the proposed approach is that it builds
the modeling from the NHPP density. By casting the nonparametric
modeling component as a density estimation problem, we can develop
flexible classes of nonparametric mixture models that allow relatively
easy prior specification and posterior simulation, and enable
modeling for multivariate mark distributions comprising both
categorical and continuous marks. Most importantly, in the context of
marked NHPPs, the methodology proposed herein provides a unified
inference framework for the joint location-mark process, the marginal
point process, and the conditional mark distribution. 
In this way, our framework offers a nice simplification of some of the 
more general models discussed in the literature, providing an easily
interpretable platform for applied inference about marked Poisson
processes. The combination of model flexibility and relative
simplicity of our approach stands in contrast to various extensions of
Gaussian process frameworks: continuous marks lead to additional
correlation function modeling or a separate mark distribution model;
it is not trivial to incorporate categorical marks; and a spatially
changing intensity surface requires complicated non-stationary spatial
correlation.

The plan for the paper is as follows. Section \ref{NHPP} presents our
general framework of model specification for the intensity function of
unmarked temporal or spatial NHPPs. Section \ref{NHPP-marked} extends
the modeling framework to general marked Poisson processes in both a
semiparametric and fully nonparametric manner. Section 
\ref{implementation} contains the necessary details for application 
of the models developed in Sections \ref{NHPP} and \ref{NHPP-marked},
including posterior simulation and inference, prior specification, and
model checking (with some of the technical details given in an Appendix).
We note that Section \ref{inference} discusses 
general methodology related to conditional inference under a DP mixture
model framework, and is thus relevant beyond the application to NHPP 
modeling. Finally, Section \ref{data-examples} illustrates the methodology
through three data examples, and Section \ref{summary} concludes with 
discussion.

%
%

\section{Mixture specification for process intensity} 
\label{NHPP}

This section outlines the various models for unmarked NHPPs which
underlie our general framework. As described in the introduction, the
ability to factor the likelihood as in (\ref{eqn:lhd}) allows for
modeling of $f(\bm{x})=$ $\lambda(\bm{x})/\Lambda_{\mc{R}}$, the process
density, independent of $\Lambda_{\mc{R}}$, the integrated process intensity. 
The Poisson assumption implies that $N$ is sufficient for 
$\Lambda_{\mc{R}}$ in the posterior distribution and, 
in Section 4, we describe standard inference under both conjugate and
reference priors for $\Lambda_{\mc{R}}$. Because the process density has
domain restricted to the observation window $\mc{R}$, we seek flexible 
models for densities with bounded support that can provide inference for 
the NHPP intensity and its functionals without relying on specific
parametric forms or asymptotic arguments.

We propose a general family of models for NHPP densities $f(\bm{x})$
built through DP mixtures $f(\bm{x};G)$ of arbitrary kernels,
$\mr{k}^{\bm{x}}(\bm{x}; \theta)$, with support on
$\mc{R}$. Specifically,
\begin{equation}\label{eqn:dpm}
f(\bm{x}; G)  = \int \mr{k}^{\bm{x}}(\bm{x}; \theta) dG(\theta),
\hspace{0.3cm} \mr{with} 
\hspace{0.3cm} \mr{k}^{\bm{x}}(\bm{x};\theta)=0~\mr{for}~\bm{x} \notin \mc{R},
\hspace{0.3cm} \mr{and} \hspace{0.3cm}  G \sim \text{DP}(\alpha,G_{0}),
\end{equation}
where $\theta$ is the (typically multi-dimensional) kernel parametrization.  
The kernel support restriction guarantees that 
$\int_\mc{R} f(\bm{x};G) d\bm{x}=1$ and hence $\Lambda_{\mc{R}}=$ 
$\int_\mc{R} \lambda(\bm{x}) d\bm{x}$. The random mixing distribution $G$ 
is assigned a DP prior \citep{Ferg1973,Anto1974} with precision parameter
$\alpha$ and base (centering) distribution $G_{0}(\cdot) \equiv$ $G_{0}(\cdot;\psi)$ 
which depends on hyperparameters $\psi$. 
For later reference, recall the DP constructive definition \citep{Sethuraman1994}
according to which the DP generates (almost surely) discrete distributions 
with a countable number of atoms drawn i.i.d. from $G_0$. The corresponding 
weights are generated using a {\it stick-breaking} mechanism based on
i.i.d. Beta$(1,\alpha)$ (a beta distribution with mean $(1 + \alpha)^{-1}$)
draws, $\{ \zeta_{s}: s=1,2,... \}$ (drawn independently of the
atoms); specifically, the first weight is equal 
to $\zeta_{1}$ and, for $l \geq 2$, the $l$-th weight is given by 
$\zeta_{l} \prod_{s=1}^{l-1} (1-\zeta_{s})$. The choice of a DP prior allows us to draw 
from the existing theory, and to utilize well-established techniques for 
simulation-based model fitting.

The remainder of this section describes options for specification of
the kernel and base distribution for the model in (\ref{eqn:dpm}): for
temporal processes in Section 2.1 and for spatial processes in Section
2.2. In full generality, NHPPs may be defined over an
unbounded space, so long as the intensity is locally integrable, but
in most applications the observation window is bounded and this will
be a characteristic of our modeling framework. Indeed, the specification 
of DP mixture models for densities with bounded support is a useful aspect
of this work in its own right. Hence, temporal point processes can be rescaled 
to the unit interval, and we will thus assume that $\mc{R}=$ $(0,1)$.
Furthermore, we assume that spatial processes are observed over rectangular 
support, such that the observation window can also be rescaled, in particular,
$\mc{R}=(0,1)\times(0,1)$ in Section \ref{spatial-NHPP} and elsewhere for 
spatial data.

\subsection{Temporal Poisson processes} 
\label{temporal-NHPP}

Denote by $\{ t_{1},\ldots,t_N \}$ the temporal point pattern observed in 
interval $\mc{R} = (0,1)$, after the rescaling described above. Following 
our factorization of the intensity as $\lambda(t)=$ $\Lambda_{\mc{R}} f(t)$ 
and conditional on $N$, the observations are assumed to arise i.i.d. from 
$f(t;G)=$ $\int \mr{k}^t(t;\theta)dG(\theta)$ and $G$ is assigned a DP prior 
as in (\ref{eqn:dpm}). We next consider specification for $\mr{k}^t(t;\theta)$.

Noting that mixtures of beta densities can approximate arbitrarily
well any continuous density defined on a bounded interval
\citep[e.g.,][Theorem 1]{DiacYlvi1985},
the beta emerges as a natural choice for the NHPP density
kernel. Therefore, the DP mixture of beta densities model for the NHPP
intensity is given by
\begin{equation}\label{beta-model}
\lambda(t;G) = \Lambda_{\mc{R}} \int b(t;\mu,\tau) dG(\mu,\tau), 
\hspace{0.3cm} t \in (0,1);
\hspace{0.5cm} G \sim \mr{DP}(\alpha,G_{0}).
\end{equation}
Here, $b(\cdot;\mu,\tau)$ denotes the density of the beta
distribution parametrized in terms of its mean $\mu \in (0,1)$ and a
scale parameter $\tau > 0$, i.e., $b(t;\mu,\tau) \propto$
$t^{\mu \tau - 1} (1-t)^{\tau (1 - \mu) - 1}$, $t \in (0,1)$.
Regarding the DP centering distribution $G_{0} \equiv$
$G_{0}(\mu,\tau)$, we work with independent components, specifically, a 
uniform distribution on $(0,1)$ for $\mu$, and an inverse gamma distribution 
for $\tau$ with fixed shape parameter $c$ and mean $\beta/(c-1)$ (provided 
$c>1$). Hence, the density of $G_{0}$ is $g_{0}(\mu,\tau) \propto$ 
$\tau^{-(c+1)}\exp(-\beta \tau^{-1})\ds{1}_{\mu \in (0,1)}$, where $\beta$ can be 
assigned an exponential hyperprior.

The beta kernel is appealing due to its flexibility and the fact that
it is directly bounded to the unit interval.  However, there are no
commonly used conjugate priors for its parameters; there are conjugate
priors for parameters of the exponential family representation of the
beta density, such as the beta-conjugate distribution in
\citet{GrunRaftGutt1993a}, but none of these are easy to work with or
intuitive to specify.  There are substantial benefits (refer to
Section 4) to be gained from the Rao-Blackwellization of posterior
inference for mixture models \citep[see, e.g.,][for empirical
demonstration of the improvement in estimators]{MacEClydLiu1999} that
is only possible with conditional conjugacy -- that is, in this
context, when the base distribution is conjugate for the kernel
parametrization. Moreover, the nonparametric mixture allows inference
to be robust to a variety of reasonable kernels, such that the
convenience of conjugacy will not usually detract from the quality of
analysis.

We are thus motivated to provide a conditionally conjugate alternative
to the beta model, and do so by first applying a logit transformation,
$\mr{logit}(t)=$ $\log\left(t/(1-t)\right)$, $t \in (0,1)$, and then using 
a Gaussian density kernel.
In detail, the logit-normal DP mixture model is then,
\begin{equation}\label{normal-model}  
\lambda(t;G) = \Lambda_{\mc{R}} 
\int \mr{N}\left(\mr{logit}(t);\mu,\sigma^2\right) \frac{1}{t(1-t)} 
dG(\mu,\sigma^{2}),  \hspace{0.3cm} t \in (0,1);
\hspace{0.5cm} G \sim \mr{DP}(\alpha,G_{0}).
\end{equation}
The base distribution is taken to be of the standard conjugate form
\citep[as in, e.g.,][]{EscoWest1995}, such that $g_{0}(\mu, \sigma^{2})=$
$\mr{N}(\mu; \delta, \sigma^2/\kappa)\mr{ga}(\sigma^{-2} ;\nu,\omega)$, 
where $\mr{ga}(\cdot;\nu,\omega)$ denotes the gamma density with 
$\ds{E}[\sigma^{-2}]=$ $\nu/\omega$. A gamma prior is placed on $\omega$
whereas $\kappa$, $\nu$ and $\delta$ are fixed (however, a normal prior 
for $\delta$ can be readily added).

The price paid for conditional conjugacy is that the logit-normal
model is susceptible to boundary effects: the density specification in
(\ref{normal-model}) {\it must} be zero in the limit as $t$ approaches
the boundaries of the observation window (such that $\mr{logit}(t)
\rightarrow \pm \infty$).  In contrast, the beta model is not
restricted to any single type of boundary behavior, and will thus be
more appropriate whenever there is a need to model processes which
maintain high intensity at the edge of the observation window.
Section 5 offers empirical comparison of the two models.

The beta and logit-normal mixtures form the basis for our approach to
modeling marked Poisson processes, and Section 2.2 will extend these
models to spatial NHPPs. Both schemes are developed to be as
flexible as possible, in accordance with our semiparametric strategy
of having point event data restricted by the Poisson assumption but
modeled with an unrestricted NHPP density. However, in some situations 
it may be of interest to constrain the model further by making structural 
assumptions about the NHPP density, including monotonicity assumptions for
the intensity function as in, for example, software reliability applications
\citep[e.g.,][]{KuoYang1996}. To model monotonic intensities for temporal 
NHPPs, we can employ the representation of non-increasing densities on 
$\ds{R}^{+}$ as scale mixtures of uniform densities. In particular, for any 
non-increasing density $h(\cdot)$ on $\ds{R}^{+}$ there exists a distribution
function $G$, with support on $\ds{R}^{+}$, such that $h(t) \equiv$
$h(t;G)=$ $\int \theta^{-1} \ds{1}_{t \in (0,\theta)}
\text{d}G(\theta)$ \citep[see, e.g.,][]{BrunLo1989,KottGelf2001}.
In the context of NHPPs, a DP mixture formulation could be written
$\lambda(t;G)=$ $\Lambda_{\mc{R}} \int \theta^{-1} \ds{1}_{t\in(0,\theta)}dG(\theta)$, 
$t \in (0,1)$, with $G \sim \mr{DP}(\alpha,G_{0})$, where $G_{0}$ has support 
on $(0,1)$, e.g., it can be defined by a beta distribution. Then, $\lambda(t;G)$ 
defines a prior model for non-increasing intensities. Similarly, a prior 
model for non-decreasing NHPP intensities can be built from 
$f(t;G)=$ $\int \theta^{-1} \ds{1}_{(t-1) \in (-\theta,0)} dG(\theta)$, $t \in (0,1)$, 
with $G \sim \mr{DP}(\alpha,G_{0})$, where again $G_0$ has support on
$(0,1)$.

\subsection{Spatial Poisson processes} 
\label{spatial-NHPP}

We now present modeling for spatial NHPPs as an extension of the framework in 
Section \ref{temporal-NHPP}. As mentioned previously, we assume that the bounded 
event data has been rescaled such that point locations $\{\bm{x}_1,\ldots,\bm{x}_N\}$ 
all lie within the unit square, $\mc{R}=$ $(0,1) \times (0,1)$. The extra implicit 
assumption of a rectangular observation window is standard in the literature on 
spatial Poisson process modeling \citep[see, e.g.,][]{Digg2003}.

The most simple extension of our models for temporal NHPPs is to build a bivariate 
kernel out of two independent densities. For example, a two-dimensional version of 
the beta mixture density in (\ref{beta-model}) could be written
$f(\bm{x};G)=$ $\int b(x_1;\mu_1,\tau_1) b(x_2;\mu_2,\tau_2)dG(\bs{\mu},\bs{\tau})$,
where $\bs{\mu}=$ $(\mu_{1},\mu_{2})$ and $\bs{\tau}=$ $(\tau_{1},\tau_{2})$. However, 
although dependence between $x_1$ and $x_2$ will be induced by mixing, it will 
typically be more efficient to allow for explicit dependence in the kernel.
A possible two-dimensional extension of (\ref{beta-model}) is that of 
\citet{KottSans2007}, which employs a Sarmanov dependence factor to induce a 
bounded bivariate density with beta marginals. The corresponding model for the 
spatial NHPP intensity is given by
\begin{eqnarray}\label{bivarbeta}
\lambda(\bm{x};G) = \Lambda_{\mc{R}} \int b(x_1; \mu_1, \tau_1) b(x_2; \mu_2,
\tau_2)\left(1+\rho(x_1-\mu_1)(x_2-\mu_2)\right)dG(\bs{\mu},\bs{\tau}, \rho),
\end{eqnarray}
where $G \sim \mr{DP}(\alpha,G_{0})$ and $G_0$ is built from independent centering 
distributions as in (\ref{beta-model}) for each dimension, multiplied by a 
conditional uniform distribution for $\rho$ over the region such that 
$1 + \rho(x_1 - \mu_1) (x_2 - \mu_2) > 0$, for all $\bm{x} \in \mc{R}$. 
Thus, $g_{0}(\bs{\mu}, \bs{\tau}, \rho)=$ $ \ds{1}_{\rho \in ( C_{\bs{\mu}}, C^{\bs{\mu}})}
(C^{\bs{\mu}}-C_{\bs{\mu}})^{-1} \prod_{i=1}^{2}
\mr{ga}(\tau^{-1}_i ; \nu_i, \beta_i ) \ds{1}_{\mu_i\in (0,1)}$,
where $C_{\bs{\mu}}=$ 
$-\left(\mr{max}\{\mu_1\mu_2,(1-\mu_1)(1-\mu_2)\}\right)^{-1}$ and
$C^{\bs{\mu}} $ $=
-\left(\mr{min}\{\mu_1(\mu_2-1),\mu_2(\mu_1-1)\}\right)^{-1}$.  Gamma
hyperpriors can be placed on $\beta_{1}$ and $\beta_{2}$.

Model (\ref{bivarbeta}) has appealing flexibility, including resistance to 
edge effects, but a lack of conditional conjugacy requires the use of an 
augmented Metropolis-Hastings algorithm for posterior simulation 
(discussed in Appendix A.2). 
The inefficiency of this approach is only confounded in
higher dimensions, and becomes especially problematic when we extend
the models to incorporate process marks. Hence, we are again motivated
to seek a conditionally conjugate alternative for spatial NHPPs, and
this is achieved in a straightforward manner by applying individual
logit transformations to each coordinate dimension and mixing over
bivariate Gaussian density kernels.
Specifically, the spatial NHPP logit-normal model is
\begin{equation}\label{bivarnormal}
  \lambda(\bm{x};G) = \Lambda_{\mc{R}} \int \mr{N}\left(\mr{logit}(\bm{x}); 
\bs{\mu},\bs{\Sigma} \right)\frac{1}{\prod_{i=1}^2 x_i (1-x_i)}
dG(\bs{\mu},\bs{\Sigma}), 
\hspace{0.5cm} G \sim \mr{DP}(\alpha,G_{0}),
\end{equation}
where $\mr{logit}(\bm{x})$ is shorthand for $\left[
 \mr{logit}(x_1), \mr{logit}(x_2) \right]'$.  The base distribution
is again of the standard conjugate form, such that 
$g_{0}(\bs{\mu},\bs{\Sigma})=$ $\mr{N}(\bs{\mu}; \bs{\delta},
\bs{\Sigma}/\kappa)\mr{W}(\bs{\Sigma}^{-1} ; \nu, \bs{\Omega})$, with fixed 
$\kappa$, $\nu$, $\bs{\delta}$ and a Wishart hyperprior for $\bs{\Omega}$.
Here, $\mr{W}(\cdot;\nu,\bs{\Omega})$ denotes a Wishart density such that
$\ds{E}[\bs{\Sigma}^{-1}]=$ $\nu \bs{\Omega}^{-1}$ and
$\ds{E}[\bs{\Sigma}]=$ $(\nu - \frac{3}{2})^{-1}\bs{\Omega}$.

%
%

\section{Frameworks for modeling marked Poisson processes} 
\label{NHPP-marked}

The models for unmarked NHPPs, as introduced in Section 2, are
essentially density estimators for distributions with bounded support.
As mentioned in the Introduction, the Bayesian nonparametric approach
is most powerful when embedded in a more complex model for marked
point processes.  Section 3.1 describes how the methodology of Section
2 can be coupled with general regression modeling for marks, whereas
in Section 3.2, we develop a fully nonparametric Bayesian modeling
framework for marked Poisson processes.

\subsection{Semiparametric modeling for the mark distribution} 
\label{marks-SP}

In the standard marked point process setting, one is interested in
inference for the process intensity over time or space and the
associated conditional distribution for the marks.

Regarding the data structure, for each temporal or spatial point $\bm{x}_{i}$, 
$i=1,...,N$, in the observation window $\mc{R}$ there is an associated mark 
$\bm{y}_{i}$ taking values in the mark space $\mc{M}$, which may be multivariate 
and may comprise both categorical and continuous variables. 
Let $h(\bm{y} \mid \bm{x})$ denote the conditional mark density at point 
$\bm{x}$. (Note that we use $\bm{y}$ and $\bm{y}_{i}$ as simplified notation 
for $\bm{y}(\bm{x})$ and $\bm{y}(\bm{x}_{i})$.)
Under the semiparametric approach, we build the joint model for the marks 
and the point process intensity through 
\begin{equation}\label{eqn:joint}
\phi(\bm{x}, \bm{y}) = \lambda(\bm{x}) h(\bm{y} \mid \bm{x}) = 
\Lambda_{\mc{R}} f(\bm{x}) h(\bm{y} \mid \bm{x}), \,\,\,\,\,
\bm{x} \in \mc{R}, \, \bm{y} \in \mc{M}.
\end{equation}
Note that the conditioning in $h(\bm{y} \mid \bm{x})$ does not involve
any portion of the point process other than point
$\bm{x}$; for instance, for temporal processes, the conditional mark
density at time $t$ does not depend on earlier times $t^{\prime} < t$.
Under this setting, the {\it Marking theorem} \citep[e.g., proposition
3.9 in][p. 55]{MollWaag2004,King1993} yields that the marked point
process $\{ (\bm{x},\bm{y}): \bm{x} \in \mc{R}, \bm{y} \in \mc{M} \}$
is a NHPP with intensity function given by (\ref{eqn:joint}) for
$(\bm{x},\bm{y}) \in$ $\mc{R} \times \mc{M}$, and by its extension to
$\mc{B} \times \mc{M}$ for any bounded $\mc{B} \supset \mc{R}$.

This intensity factorization, combined with the general NHPP
likelihood factorization in (\ref{eqn:lhd}), results in convenient
semiparametric modeling formulations for the marked process through a
DP mixture model for $f(\cdot)$ (as in Section \ref{NHPP}) and a
separate parametric or semiparametric regression specification for the
conditional mark distribution.  In particular, assuming that the marks
$\{\bm{y}_i\}_{i=1}^N$ are mutually independent given
$\{\bm{x}_i\}_{i=1}^N$, and combining (\ref{eqn:lhd}) and
(\ref{eqn:joint}), we obtain
\begin{equation}\label{markslhd}
  \mr{p}\left(\{\bm{x}_i,\bm{y}_i\}_{i=1}^{N} ;\Lambda_{\mc{R}},f(\cdot),h(\cdot) \right)
  \propto \Lambda_{\mc{R}}^{N} \exp(-\Lambda_{\mc{R}})\prod_{i=1}^{N} f(\bm{x}_i)
  \prod_{i=1}^{N} h(\bm{y}_i \mid \bm{x}_i),
\end{equation}
such that the conditional mark density can be modeled independent
of the process intensity.

The consequence of this factorization of integrated intensity, process
density, and the conditional mark density, is that any regression
model for $h$ can be added onto the modeling schemes of Section
\ref{NHPP} and provide an extension to marked processes. In some
applications, it will be desirable to use flexible semiparametric
specifications for $h$, such as a Gaussian process regression model,
while in other settings it will be useful to fit $h$ parametrically,
such as through the use of a generalized linear model. As an
illustration, Section \ref{simulation} explores a Gaussian
process-based specification, however, the important point is that this
aspect of the modeling does not require any further development of the
underlying nonparametric model for the NHPP intensity. Moreover,
despite the posterior independence of $f$ and $h$, combining them as
in (\ref{eqn:joint}) leads to a practical semiparametric inference
framework for the joint mark-location Poisson process.  The fully
nonparametric approach developed in the following section provides an
alternative for settings where further modeling flexibility is needed.

\subsection{Fully nonparametric joint and implied conditional mark modeling} 
\label{marks-NP}

While the semiparametric approach of Section 3.1 provides a convenient
extension of the NHPP models in Section 2, the connection
between joint and marked processes provides the opportunity to build
fully nonparametric models for marked point event data. Here, we introduce 
a general modeling approach, built through fully nonparametric models for 
joint mark-location Poisson processes, and describe how this provides a 
unified inference framework for the joint process, the conditional mark 
distribution, and the marginal point process.

Instead of specifying directly a model for the marked process, we begin by 
writing the joint Poisson process, PoP$(\mc{R} \times \mc{M},\phi)$, defined 
over the joint location-mark observation window with intensity 
$\phi(\bm{x},\bm{y})$. The inverse of the marking theorem used to obtain 
equation (\ref{eqn:joint}) holds that, if the marginal intensity 
$\int_\mc{M} \phi(\bm{x},\bm{y})d\bm{y}=$ $\lambda(\bm{x})$ is locally 
integrable, then the joint process just defined is also the marked Poisson 
process of interest.

Analogously to the model development in Section 2, we define a process
over the joint location-mark space with intensity function
\begin{eqnarray} 
\phi(\bm{x},\bm{y};G) = 
\Lambda_{\mc{R}} \int  \mr{k}^{\bm{x}}(\bm{x} ; \theta^{\bm{x}}) 
\mr{k}^{\bm{y}}(\bm{y} ; \theta^{\bm{y}}) dG(\theta^{\bm{x}}, \theta^{\bm{y}}) = 
\Lambda_{\mc{R}} f(\bm{x},\bm{y};G), 
\hspace{0.5cm} G \sim \mr{DP}(\alpha,G_{0}),
\label{lmproc}
\end{eqnarray} 
where the mark kernel $\mr{k}^{\bm{y}}(\bm{y}; \theta^{\bm{y}})$ has
support on $\mc{M}$ and the integrated intensity can be defined in
terms of either the joint or marginal process, such that
$\Lambda_{\mc{R}}=$ $\int_\mc{R} \lambda(\bm{x}) d\bm{x}=$ 
$\int_\mc{R} \left[\int_\mc{M} \phi(\bm{x},\bm{y}) d\bm{y} \right]d\bm{x}$.  
Note that the marginal intensity, and hence the marked point process, are
properly defined with locally integrable intensity functions.
Specifically, we can move integration over $\mc{M}$ inside the infinite
sum and
\begin{eqnarray} 
\int_\mc{M} \phi(\bm{x}, \bm{y}) d\bm{y} & = & \Lambda_{\mc{R}}
\int_{\theta_\bm{x}} \mr{k}^{\bm{x}}(\bm{x} ; \theta^{\bm{x}}) \int_{\theta_\bm{y}}  
\left[ \int_\mc{M} \mr{k}^{\bm{y}}(\bm{y} ; \theta^{\bm{y}}) d\bm{y} \right]
dG(\theta^{\bm{x}}, \theta^{\bm{y}})  \label{eqn:locint} \\ 
 & = & \Lambda_{\mc{R}} 
\int \mr{k}^{\bm{x}}(\bm{x};\theta^{\bm{x}})dG^{\bm{x}}(\theta^{\bm{x}}) =
\Lambda_{\mc{R}} f({\bm{x}}; G) = \lambda(\bm{x}). \notag
\end{eqnarray}
Here, $G^{\bm{x}}(\theta^{\bm{x}})$ is the marginal mixing distribution, which 
has an implied DP prior with base density $g_{0}^{\bm{x}}(\theta^{\bm{x}})=$ 
$\int g_{0}(\theta^{\bm{x}},\theta^{\bm{y}}) d\theta^{\bm{y}}$, and we have
thus recovered the original DP mixture model of Section 2 for the
marginal location NHPP
$\mathrm{PoP}(\mc{R},\lambda)$. As an aside we note that, through a
similar argument and since $\phi(\bm{x},\bm{y})=$
$\lambda(\bm{x}) h(\bm{y} \mid \bm{x})$, the joint location-mark
process of (\ref{lmproc}) satisfies the requirements of proposition
3.9 in \cite{MollWaag2004}, and hence the marks alone are marginally
distributed as a Poisson process defined on $\mc{M}$ with intensity 
$\int_{\mc{R}} \phi(\bm{x},\bm{y}) d\bm{x}=$ $\Lambda_{\mc{R}} 
\int \mr{k}^{\bm{y}}(\bm{y};\theta^{\bm{y}})dG^{\bm{y}}(\theta^{\bm{y}})$.

In general, both the mixture kernel and base distributions will be built
from independent components corresponding to marks and to locations, and 
the random mixing measure is relied upon to induce dependence between 
these random variables. This technique has been employed
in regression settings by \cite{TaddKott2010}, and provides a fairly 
automatic procedure for nonparametric model building in mixed data-type 
settings. For example, suppose that a spatial point process is accompanied 
by categorical marks, such that marks $\{y_1,\ldots,y_N\}$ are each a member 
of the set $\mc{M}=$ $\{1,2,\ldots,\mr{M}\}$. The joint intensity model 
can be specified as
\begin{equation}\label{catmodel}
\phi(\bm{x},y;G) = \Lambda_{\mc{R}} \int \mr{k}^\bm{x}(\bm{x};\theta^{\bm{x}})
q_{y} dG(\theta^{\bm{x}},\bm{q}),
\hspace{0.5cm} 
G \sim \mr{DP}(\alpha,G_{0}^{\bm{x}}(\theta^{\bm{x}})\mr{Dir}(\bm{q};\bm{a})),
\end{equation}
where $\bm{q} = [q_1,\ldots,q_{\mr{M}}]$ is a probability vector with
$q_y=$ $\mr{Pr}(Y=y \mid \bm{q})$, $\mr{Dir}(\bm{q};\bm{a})$ is the
Dirichlet distribution, with $\bm{a}=$ $(a_{1},...,a_{M})$, such that 
$\mathbb{E}(q_y \mid \bm{a})=$
$a_y/\sum_{s=1}^{\mr{M}} a_s$, and the location-specific kernel, $\mr{k}^{\bm{x}}$,
and centering distribution, $G_{0}^{\bm{x}}$, are specified as in either 
(\ref{bivarbeta}) or (\ref{bivarnormal}) and thereafter. Additional marks 
can be incorporated in the same manner by including additional independent 
kernel and base distribution components.

Similarly, continuous marks can be modeled through an appropriate
choice for the independent mark kernel. For example, in the case of
real-valued continuous marks (i.e., $\mc{M}=\mathbb{R}$) for a temporal
point process, the choice of a normal density kernel leads to the
intensity model
\begin{eqnarray}\label{ctsmodel}
\phi(t,y;G) = \Lambda_{\mc{R}} \int \mr{k}^t(t;\theta^t) \mr{N}(y;\eta,\sigma^2) 
dG(\theta^t,\eta,\sigma^{2}), \,\,\,
G \sim \mr{DP}\left(\alpha,G_{0}^{t}(\theta^{t})G_{0}^{y}(\eta,\sigma^{2}) \right).
\end{eqnarray}
The location specific kernel, $\mr{k}^t$, and base measure,
$G_{0}^{t}$, can be taken from Section \ref{temporal-NHPP};
$G_{0}^{y}$ can be specified through the conjugate normal inverse-gamma form 
as in (\ref{normal-model}). Other possible mark kernels are negative-binomial 
or Poisson for count data (as in Section \ref{coal-mining}), a Weibull for 
failure time data, or a log-normal for positive continuous marks (as in 
Section \ref{pine-trees}).

As an alternative to this generic independent kernel approach, the
special case of a combination of real-valued continuous marks with the
logit-normal kernel models in either (\ref{normal-model}) or
(\ref{bivarnormal}) allows for joint multivariate-normal kernels.
Thus, instead of the model in (\ref{ctsmodel}), a temporal point process 
with continuous marks is specified via bivariate normal kernels as
\begin{equation}\label{mvnmarks}
  \phi(t,y;G) = \Lambda_{\mc{R}} \int \mr{N}\left([\mr{logit}(t),y]'; 
\bs{\mu},\bs{\Sigma} \right)\frac{1}{t(1-t)} d
  G(\bs{\mu},\bs{\Sigma}), 
  \hspace{0.5cm} G \sim \mr{DP}(\alpha,G_{0}),
\end{equation}
with base distribution of the standard conjugate form, exactly as
described following (\ref{bivarnormal}). Specification is easily
adapted to spatial processes or multivariate continuous marks through
the use of higher dimensional normal kernels (see Section \ref{pine-trees}
for an illustration).

A key feature of the joint mixture modeling framework for the location-mark
process is that it can provide flexible specifications for multivariate 
mark distributions comprising both categorical and continuous marks.
For any of the joint intensity models specified in this section, inference 
for the conditional mark density is available through
\begin{equation}\label{markcnd}
h(\bm{y} \mid \bm{x} ; G) =
 \frac{f(\bm{x},\bm{y} ; G)}{f(\bm{x} ; G)} = 
\frac{\int \mr{k}^{\bm{x}}(\bm{x} ; \theta^{\bm{x}}) 
\mr{k}^{\bm{y}}(\bm{y} ; \theta^{\bm{y}})  dG(\theta^{\bm{x}}, \theta^{\bm{y}}) }
{\int \mr{k}^{\bm{x}}(\bm{x} ; \theta^{\bm{x}}) dG^{\bm{x}}(\theta^{\bm{x}}) }.
\end{equation}
Of course, other conditioning arguments are also possible if, for example, 
some subset of the marks is viewed as covariates for a specific mark of 
interest. In any case, the integrals in (\ref{markcnd}) are actually infinite 
sums induced by discrete realizations from the posterior distribution for $G$. 
%
%
In Section \ref{inference}, we show that truncation approximations to the 
infinite sums allow for proper conditional inference and, hence, for fully 
nonparametric 
inference about any functional of the conditional mark distribution.

%
%

\section{Implementation} 
\label{implementation}

This section provides guidelines for application of the models
proposed in Sections \ref{NHPP} and \ref{NHPP-marked}, with prior
specification and posterior simulation briefly discussed in Section
\ref{MCMC} (further details can be found in the Appendix), inference 
for marked NHPP functionals in Section \ref{inference}, and model 
checking in Section \ref{model-checking}.

\subsection{Prior specification and posterior simulation} 
\label{MCMC}

As with our approach to model building, we can specify the prior for
integrated intensity independent of the prior for parameters of the DP
mixture density model. The marginal likelihood for $\Lambda_{\mc{R}}$
corresponds to a Poisson density for $N$, such that the conjugate
prior for $\Lambda_{\mc{R}}$ is a gamma distribution. As a default
alternative, we make use of the (improper) reference prior for
$\Lambda_{\mc{R}}$, which can be derived as $\pi(\Lambda_{\mc{R}})
\propto \Lambda_{\mc{R}}^{-1}$ for $\Lambda_{\mc{R}} > 0$.  The
posterior distribution for the integrated intensity is then
available analytically as a gamma distribution, since the posterior
distribution for the NHPP intensity factorizes as
$\mr{p}(f(\cdot),\Lambda_{\mc{R}} \mid \mr{data})=$ $\mr{p}(f(\cdot)
\mid \mr{data}) \mr{p}(\Lambda_{\mc{R}} \mid N)$.  In particular,
$\mr{p}(\Lambda_{\mc{R}} \mid N)=$ $\mr{ga}(N,1)$ under our default
reference prior. Similarly, under the semiparametric
approach of Section \ref{marks-SP}, prior specification and posterior
inference for any model applied to the conditional mark distribution
can be dealt with separately from the intensity function model, and will
generally draw on existing techniques for the regression model of
interest.

What remains is to establish general prior specification and MCMC
simulation algorithms for the DP mixture process density models of
Sections \ref{NHPP} and \ref{marks-NP}.  In a major benefit of our
approach -- one which should facilitate application of these models --
we are able here to make use of standard results and methodology from
the large literature on DP mixture models.  Our practical
implementation guidelines are detailed in the Appendix, with prior
specification in A.1 and a posterior simulation framework in A.2.

\subsection{Inference about NHPP functionals} 
\label{inference}

Here, we describe the methods for posterior inference about joint or 
marginal intensity functions and for conditional density functions. 
We outline inference for a general NHPP with events $\{\bm{z}_i\}_{i=1}^N$, 
possibly consisting of both point location and marks, and leave specifics 
to the examples of Section \ref{data-examples}.

Due to the almost sure discreteness of the DP, a generic
representation for the various mixture models for NHPP densities is
given by $f(\bm{z};G)=$ $\sum_{l=1}^{\infty} p_{l} \mr{k}(\bm{z};
\vartheta_{l})$, where the $\vartheta_{l}$, given the base
distribution hyperparameters $\psi$, are i.i.d. from $G_{0}$, and the
weights $p_{l}$ are generated according to the stick-breaking process
discussed in Section \ref{NHPP}. Here, $\bm{z}$ may include only point
locations (as in the models of Section \ref{NHPP}) or both point
locations and marks whence $\mr{k}(\bm{z}; \vartheta)=$
$\mr{k}^{\bm{x}}(\bm{x};\vartheta^{\bm{x}}) \mr{k}^{\bm{y}}(\bm{y} ;
\vartheta^{\bm{y}})$ (as in Section \ref{marks-NP}).  Hence, the DP
induces a clustering of observations: for $\text{data}=$
$\{\bm{z}_1,\ldots,\bm{z}_N\}$, if we introduce latent mixing
parameters $\bs{\theta}=$ $\{\theta_1,\ldots,\theta_N\}$ such that
$\bm{z}_i \mid \theta_{i} \stackrel{ind}{\sim}$
$\mr{k}(\bm{z}_i;\theta_i)$, with $\theta_i \mid G
\stackrel{iid}{\sim} G$, for $i=1,\ldots,N$, and $G \mid \alpha,\psi
\sim$ DP$(\alpha,G_{0}(\cdot;\psi))$, then observations can be grouped
according to the number, $m \le N$, of distinct mixing parameters in
$\bs{\theta}$.  This group of distinct parameter sets,
$\bs{\theta}^{\star} =$ $\{\theta^\star_1,\ldots,\theta_m^\star\}$,
maps back to data through the latent allocation vector, $\bm{s}=$
$[s_1,\ldots,s_N]$, such that $\theta_i=$ $\theta^{\star}_{s_i}$. The
expanded parametrization is completed by the number of observations
allocated to each unique component, $\bm{n}=$ $[n_1,\ldots,n_m]$,
where $n_j=$ $\sum_{i=1}^{N} \ds{1}_{s_i=j}$, and the associated
groups of observations $\{\bm{z}_i: s_i=j\}$.  If $G$ is marginalized
over its DP prior, we obtain the P\'olya urn expression for the DP
prior predictive distribution,
\begin{equation}\label{polya}
\mr{p}(\theta_{0} \mid \bs{\theta}^\star,\alpha,\psi) = 
d\ds{E}\left[G(\theta_{0}) \mid \bs{\theta}^{\star},\alpha,\psi \right] \propto 
\alpha g_{0}(\theta_{0};\psi) + \sum_{j=1}^{m} n_{j} \delta_{\theta^{\star}_j}(\theta_{0})
\end{equation}
where $\delta_{a}$ denotes a point mass at $a$. Moreover, based on the DP P\'olya 
urn structure, the prior for $\bs{\theta}^\star$, given $m$ and $\psi$, is such 
that $\theta^{\star}_{j} \mid \psi \stackrel{iid}{\sim} G_{0}(\cdot;\psi)$, for 
$j=1,\ldots,m$.

Within the DP mixture framework, estimation of linear functionals of
the mixture is possible via posterior expectations conditional on only
this finite dimensional representation (i.e., it is not
necessary to draw $G$). In particular, with the NHPP density modeled
as our generic DP mixture, the posterior expectation for the intensity
function can be written as
%
%
$\ds{E}\left[\lambda(\bm{z};G) \mid \text{data} \right]=$
$\ds{E}(\Lambda_{\mc{R}} \mid N) \mr{p}(\bm{z} \mid \text{data})$,
where $\mr{p}(\bm{z} \mid \text{data})=$ 
$\ds{E}\left[ f(\bm{z};G) \mid \text{data} \right]$ is the posterior 
predictive density given by 
\begin{equation}\label{expect} 
\int  \frac{1}{\alpha+N}
\left(
\alpha  \int \mr{k}(\bm{z};\theta) dG_0(\theta;\psi) 
+ \sum_{j=1}^{m} n_{j} \mr{k}(\bm{z};\theta^{\star}_{j}) 
\right) \mr{p}(\bs{\theta}^{\star},\bm{s},\alpha,\psi \mid \text{data})
d\bs{\theta}^{\star} d\bm{s} d\alpha d\psi.
\end{equation}
Hence, a point estimate for the intensity function is readily available
through $\ds{E}\left[ f(\bm{z};G) \mid \text{data} \right]$ estimated
as the average, for each point in a grid in $\bm{z}$, over realizations of
(\ref{expect}) calculated for each MCMC posterior sample for
$\bs{\theta}^{\star}$, $\bm{s}$, $\alpha$ and $\psi$.

However, care must be taken when moving to posterior inference about the
conditional mark distribution in (\ref{markcnd}). As a general point
on conditioning in DP mixture models for joint distributions,
P\'{o}lya urn-based posterior expectation calculations, such as (\ref{expect}), 
are invalid for the estimation of non-linear functionals of $\lambda$ or $f$. 
For example, \citet{MullErkaWest1996} develop a DP mixture curve fitting 
approach that, in the context of our model, would estimate the conditional 
mark density by 
\begin{equation}\label{muller}
  \widehat{h(\bm{y} | \bm{x} )} =
  \int \frac{ \int 
    \mr{k}^{\bm{x}}(\bm{x} ; \theta^{\bm{x}}) 
\mr{k}^{\bm{y}}(\bm{y} ; \theta^{\bm{y}}) 
d\ds{E}\left[G(\theta ) \mid \bs{\theta}, \alpha, \psi\right]}{\int 
    \mr{k}^{\bm{x}}(\bm{x};\theta^{\bm{x}})
d\ds{E}\left[G(\theta ) \mid \bs{\theta}, \alpha, \psi\right]}
\mr{p}(\bs{\theta}, \alpha,\psi \mid \mr{data})
d\bs{\theta} d\alpha d\psi,
\end{equation}
which is the ratio of P\'{o}lya urn joint and marginal density point 
estimates given $\bs{\theta}$ and DP prior parameters $\alpha$, $\psi$,
averaged over MCMC draws for these parameters.  Unfortunately,
(\ref{muller}) is {\it not} 
$\mathbb{E}\left[ h(\bm{y} \mid \bm{x};G) \mid \text{data} \right]$, 
the posterior expectation for random conditional density 
$h(\bm{y} \mid \bm{x};G)=$
$f(\bm{x},\bm{y};G)/f(\bm{x};G)$, which would be the natural estimate
for the conditional mark density at any specified combination of
values $(\bm{x},\bm{y})$. Hence, the regression estimate in 
\citet{MullErkaWest1996} as well as that proposed in the more recent work
of \citet{RodrDunsGelf2009}, based on 
$\mr{p}(\bm{x},\bm{y} \mid \text{data})/\mr{p}(\bm{x} \mid \text{data})$,
provide only approximations to 
$\mathbb{E}\left[ h(\bm{y} \mid \bm{x};G) \mid \text{data} \right]$; in 
particular, the latter estimate is approximating the expectation of a ratio 
with the ratio of expectations. 
Such approximations become particularly difficult to justify if one seeks
inference for non-linear functionals of $h(\bm{y} \mid \bm{x};G)$.

Hence, to obtain the exact point estimate 
$\mathbb{E}\left[ h(\bm{y} \mid \bm{x};G) \mid \text{data} \right]$, and, most 
importantly, to quantify full posterior uncertainty about general functionals 
of the NHPP intensity, it is necessary to obtain posterior samples for the 
mixing distribution, $G$. Note that 
$\mr{p}(G \mid \text{data})=$ 
$\int \mr{p}(G \mid \bs{\theta}^{\star},\bm{s},\alpha,\psi) 
\mr{p}(\bs{\theta}^{\star},\bm{s},\alpha,\psi \mid \text{data})
d\bs{\theta}^{\star} d\bm{s} d\alpha d\psi$, where
$\mr{p}(G \mid \bs{\theta}^{\star},\bm{s},\alpha,\psi)$ follows a DP 
distribution with precision parameter $\alpha + N$ and base distribution 
given by (\ref{polya}) (see Appendix A.2).
As discussed in \cite{IshwZare2002}, using results from \cite{Pitman1996},
a draw for $G \mid \bs{\theta}^{\star},\bm{s},\alpha,\psi$
can be represented as 
$q_{0} G^{*}(\cdot)$ + $\sum_{j=1}^{m} q_{j} \delta_{\theta^{\star}_{j}}(\cdot)$,
where $G^{*} \mid \alpha,\psi \sim$ DP$(\alpha,G_{0}(\psi))$, and, 
independently of $G^{*}$, $(q_{0},q_1,...,q_{m}) \mid \alpha,\bm{s} \sim$ 
$\mr{Dir}(q_{0},q_1,...,q_{m};\alpha, n_{1},...,n_{m})$.
Therefore, posterior realizations for $G$ can be efficiently generated, by 
drawing for each posterior sample $\{ \bs{\theta}^{\star},\bm{s},\alpha,\psi \}$,
\[
dG_{L} = q_{0} \left\{ \sum_{l=1}^{L} p_{l} \delta_{\vartheta_{l}}(\cdot) \right\}
+ \sum_{j=1}^{m} q_{j} \delta_{\theta^{\star}_{j}}(\cdot),
\]
that is, using a truncation approximation to $G^{*}$ based on the DP 
stick-breaking definition. Specifically, the $\vartheta_{l}$,
$l=1,...,L$, are i.i.d. from $G_{0}(\psi)$, and 
the $p_{l}$ are constructed through i.i.d. Beta$(1,\alpha)$ draws, 
$\zeta_{s}$, $s=1,...,L-1$, such that $p_{1}=\zeta_{1}$, 
$p_{l}=$ $\zeta_{l} \prod_{s=1}^{l-1} (1-\zeta_{s})$, for
$l=2,...,L-1$, and $p_{L}=$ $1 - \sum_{l=1}^{L-1} p_{l}$. The
truncation level $L$ can be chosen using standard distributional
properties for the weights in the DP representation for $G^{*}=$
$\sum_{l=1}^{\infty} \omega_{l} \delta_{\vartheta_{l}}(\cdot)$. For instance, 
$\ds{E}(\sum\nolimits_{l=1}^{L} \omega_{l} \mid \alpha)=$
$1 - \{ \alpha/(\alpha + 1) \}^{L}$, which can be averaged over the prior
for $\alpha$ to estimate $\ds{E}(\sum\nolimits_{l=1}^{L} \omega_{l})$. Given any 
specified tolerance level for the approximation, this expression
yields the corresponding value $L$. Note that even for dispersed
priors for $\alpha$, relatively small values for $L$ (i.e., around 50)
will generally provide very accurate truncation approximations.

Now, the posterior distribution for any functional (linear or non-linear) 
of the NHPP density, and thus of the intensity function, can be sampled by 
evaluating the functional using the posterior realizations $G_{L}$. For 
example, suppose that $\bm{z}=$ $[t,y]$, such that we have a temporal process 
with a single mark, where the mixture kernel factors as $\mr{k}(\bm{z};\theta)=$
$\mr{k}^{t}(t;\theta^t)\mr{k}^y(y;\theta^y)$. Given a posterior realization
for $G_L$ and a posterior draw for $\Lambda_{\mc{R}}$, 
a posterior realization for the marginal process intensity at time $t$ is 
available as 
\[
\lambda(t;G_L) =
\Lambda_{\mc{R}} \left[ q_0 \sum\nolimits_{l=1}^L p_l \mr{k}^t(t;\vartheta^t_l) 
+ \sum\nolimits_{j=1}^m q_j \mr{k}^t(t;\theta^{\star t}_j)\right]
\]
where $\vartheta_{l}=$ $(\vartheta_{l}^{t},\vartheta_{l}^{y})$ and 
$\theta_{j}^{\star}=$ $(\theta_{j}^{\star t},\theta_{j}^{\star y})$, and a realization 
for the conditional density of mark value $y$ at time $t$ arises through
\begin{equation}\label{approxcnd}
h(y \mid t;G_L) =  \frac{ q_0 \sum_{l=1}^L p_l
\mr{k}^t(t;\vartheta^t_l) \mr{k}^y(y;\vartheta^y_l) + 
\sum_{j=1}^m q_j \mr{k}^t(t;\theta^{\star t}_j) \mr{k}^y(y;\theta^{\star y}_j)}
{q_0 \sum_{l=1}^L p_l
\mr{k}^t(t;\vartheta^t_l) + \sum_{j=1}^m q_j \mr{k}^t(t;\theta^{\star t}_j)}.
\end{equation}
Similarly, realized conditional expectation is available as 
\begin{equation}\label{cond-mean}
\ds{E}[y\mid t; G_L] = (f(t ; G_L))^{-1}
\left\{
q_0 \sum\nolimits_{l=1}^L p_l \mr{k}^t(t;\vartheta^t_l) \ds{E}(y \mid \vartheta^y_l) 
+ \sum\nolimits_{j=1}^m q_j \mr{k}^t(t;\theta^{\star t}_j) 
\ds{E}(y \mid \theta^{\star y}_j)
\right\}
\end{equation}
a weighted average of kernel means with time-dependent weights.
For multivariate Gaussian kernels, as in (\ref{mvnmarks}), one would use
conditional kernel means (available through standard multivariate normal
theory; see Section \ref{coal-mining}). The approach applies
similarly to multivariate marks and/or to marked spatial NHPP, and we
can thus obtain flexible inference for general functionals of marked
NHPPs with full uncertainty quantification. 
%
%

\subsection{Model checking} 
\label{model-checking}

A basic assumption implied by the Poisson process model is that the
number of events within any subregion of the observation window are
Poisson distributed, with mean equal to the integrated intensity over 
that subregion. Hence, a standard approach to assessing model validity is
to compare observed counts to integrated intensity within a set of
(possibly overlapping) subregions \citep[e.g.,][]{Digg2003,BTMH2005}.

An alternative approach to model checking is to look at
goodness-of-fit for simplifying transformations of the observations.
In particular, we propose transforming each margin of the point event
data (i.e., each spatial coordinate and each mark) into quantities
that are assumed, conditional on the intensity model, distributed as
i.i.d. uniform random variables. Posterior samples of these (assumed)
i.i.d. uniform sets can be compared, either graphically or formally, 
to the uniform distribution to provide a measure of model validity.

Consider first temporal point processes, and assume that the point 
pattern $\{ t_{i}: i=1,...,N \}$, with ordered time points
$0 = t_0 < t_{1} \leq t_{2} \leq ... \leq t_{N} <
1$, is a realization from a NHPP with intensity function
$\lambda(t)$ and cumulative intensity function
$\Lambda(t)=$ $\int_{0}^{t} \lambda(s) ds$.  
Then, based
on the time-rescaling theorem \citep[e.g.,][]{DaleVere2003}, the
transformed point pattern $\{ \Lambda(t_{i}): i=1,...,N \}$ is a
realization from a homogeneous Poisson process with unit rate. Let
$\Lambda(t;G_{L})$ be the posterior draws for the cumulative intensity,
obtained following the approach of Section \ref{inference}.
Then, with $\Lambda(0;G_L) = 0$ by definition, the rescaled times
$\Lambda(t_{i};G_L) - \Lambda(t_{i-1};G_L)$, $i=1,...,N$, are
independent exponential random variables with mean one. Thus, the
sampled $u_{i}=$ $1 - \exp \{ - ( \Lambda(t_{i};G_L) -
  \Lambda(t_{i-1};G_L) ) \}$, $i=1,...,N$, are independent
uniform random variables on $(0,1)$.

This approach can be extended to spatial processes by applying the
rescaling to each margin of the observation window
\citep[e.g.,][]{Cres1993}. If we have data corresponding to a NHPP on
$\mc{R}=(0,1)\times (0,1)$ with intensity $\lambda(\bm{x})$, then
point event locations along (say) the first margin of the window are
the realization of a one-dimensional NHPP with intensity
$\lambda_{1}(x_{1})=$ $\int_{0}^{1} \lambda(\bm{x})dx_2$, and
analogously for $\lambda_{2}(x_2)$. Since the kernels in
(\ref{bivarbeta}) and (\ref{bivarnormal}) are easily marginalized,
cumulative intensities $\Lambda_1(\cdot)$ and $\Lambda_2(\cdot)$ are
straightforward to calculate as sums of marginal kernel distribution
functions, based on the sampled $G_L$ as described in Section 4.2.
For each dimension $j$, these are then applied to ordered marginals
$\{x_{j,1},\ldots,x_{j,N}\}$ to obtain i.i.d. uniform random
variables, $u_{ij}=$ $1 - \exp \{ - (\Lambda_j(x_{j,i};G_L) -
\Lambda_j(x_{j,i-1};G_L)) \}$, $i=1,...,N$, where by definition
$\Lambda_j(x_{j,0};G_L) = 0$ for $j=1,2$.

Finally, there are a variety of ways that the marks can be transformed
into uniform random variables (for instance, the marginal process for
continuous marks is also Poisson, such that the time-rescaling theorem
applies), but, arguably, the most informative approach is to look at
the conditional mark distribution of (\ref{markcnd}). Full inference
is available for the conditional cumulative distribution function $H(y
\mid \bm{x};G_{L})=$ $\int_{-\infty}^{y} h(s \mid \bm{x}; G_L)ds$,
through a summation similar to that in (\ref{approxcnd}), at any
desired points $(\bm{x},y)$.  We thus obtain sets of $u_i$ that are
assumed to be i.i.d. uniform by taking, for each sampled $G_L$, the
distribution function evaluated at the data such that $u_i=$ $H(y_i
\mid \bm{x}_i;G_L)$, for $i=1,\ldots,N$.

Goodness-of-fit is evaluated through comparison of the $u_{i}$ samples 
with the uniform distribution, using either graphical or distance-based
techniques. For instance, in the context of neuronal data analysis,
\citet{BrowBarbVentKassFran2001} used standard tests and
quantile-quantile (Q-Q) plots to measure agreement of the estimated
$u_{i}$ with the uniform distribution on $(0,1)$.  
In the examples of Section \ref{data-examples}, we focus on 
Q-Q plots for graphical model assessment,
and find that these provide an intuitive picture of the marginal fit. 
In particular, under our Bayesian modeling approach, inference about model
validity can be based on samples from the full posterior for each set of 
$u_{i}$, with each realization corresponding to a single draw for $G_L$, 
through plots of posterior means and uncertainty bounds for the Q-Q graphs.

The rescaling diagnostics involve a checking of the fit provided by
the DP mixture model as well as of the Poisson process model
assumption, and thus characterize a general nonparametric model
assessment technique. Note that, in evaluating the model for 
event-location intensity, it is not, in general, feasible under this
approach to distinguish the role of the Poisson assumption from the 
form of the nonparametric model for the NHPP density. The flexibility 
of the DP mixture modeling framework is useful in this regard, since
by allowing general intensity shapes to be uncovered by the data, it
enables focusing the goodness-of-fit evaluation on the NHPP assumption
for the point process. Furthermore, all of these goodness-of-fit
assessments are focused on model validity with respect to marginal 
processes (although, of course, these are implied marginals from a 
multidimensional fit). It is possible to extend the rescaling
approach to higher dimensions, by defining a distance metric in the 
higher dimensional space and evaluating cumulative intensity functions 
with respect to this metric \citep[e.g.,][]{Diggle1990}.
However, such procedures are considerably more difficult to implement 
and will need to be designed specifically for the application of interest.

%
%

\section{Examples} 
\label{data-examples}

We include three data examples to illustrate the methodology.
Specifically, Section \ref{simulation} involves a simulated data set
from a one-dimensional Poisson process with both categorical and 
continuous marks. In Sections \ref{coal-mining} and \ref{pine-trees}, 
we consider real data on coal mining disaster events occurring in 
time with count marks, and on spatial tree locations data with 
trunk-diameter marks, respectively.

\subsection{Simulated events with continuous and binary marks}
\label{simulation}

We first consider a simulated data set from a temporal Poisson process
with observation window $\mc{R}=(0,1)$ and intensity $\lambda(t)=$
$250 \left( b(t ;1/11, 11) + b(t;4/7,7) \right)$, such that
$\Lambda_{\mc{R}} = 500$. The simulated point pattern comprises 
$N=481$ point events, which are accompanied by binary marks $z$ and
continuous marks $y$ generated from a joint conditional density 
$h(y,z \mid t)=$ $h(y \mid z,t)\mr{Pr}(z \mid t)$. Here,
$\mr{Pr}(z=1 \mid t) = t^2$ and the conditional distribution for $y$,
given $z$ and $t$, is built from $y=$ $-10(1-t)^4 + \varepsilon$, with 
$\varepsilon \sim \mr{N}(0,1)$ if $z=0$, and $\varepsilon \sim \mr{ga}(4,1)$ 
if $z=1$. Hence, the marginal regression function for $y$ given $t$ is 
non-linear with non-constant error variance, and $\mr{Pr}(z=1 \mid t)$ 
increases from 0 to 1 over $\mc{R}$.

We consider a fully nonparametric DP mixture model consisting of the 
beta kernel in (\ref{beta-model}) for point locations combined with a 
normal kernel for $y$ and a Bernoulli kernel for $z$. Hence, the full 
model for the NHPP density is given by
\[
f(t,y,z;G) = \int b(t;\mu,\tau) \mr{N}(y;\eta,\phi) q^z(1-q)^{1-z} 
dG(\mu, \tau,\eta, \phi, q), \,\,\, G \sim \mr{DP}(\alpha,G_0)
\]
where $g_{0}(\mu,\tau,\eta,\phi,q)=$ $\ds{1}_{\mu \in (0,1)}
\mr{ga}(\tau^{-1};2,\beta_{\tau}) \mr{N}(\eta;0,20\phi)
\mr{ga}(\phi^{-1};2,\beta_{\phi}) b(q;0.5,1)$.  We use the reference
prior for $\Lambda_{\mc{R}}$, and for the DP hyperpriors take $\alpha
\sim \mr{ga}(2,1)$, $\beta_{\tau} \sim \mr{ga}(1, 1/20)$ and
$\beta_{\phi} \sim \mr{ga}(1,1)$; note that $\beta_{\tau}$ and
$\beta_{\phi}$ are the means for $\tau$ and $\phi$, respectively,
under $G_0$. 
The hyperpriors are specified following the guidelines of Appendix 
A.1, and posterior simulation proceeds as outlined in Appendix A.2.
%
%
Since the beta kernel specification is non-conjugate, we jointly sample
parameters and allocation variables with Metropolis-Hasting draws for 
each $(\mu_i, \tau_i)$ and $s_i$ given $\bm{s}^{(-i)}$ and
$(\bs{\mu}^\star,\bs{\tau}^\star)^{(-i)}$, as in algorithm 5 of
\citet{Neal2000}.

\begin{figure}
\includegraphics[width=6.25in]{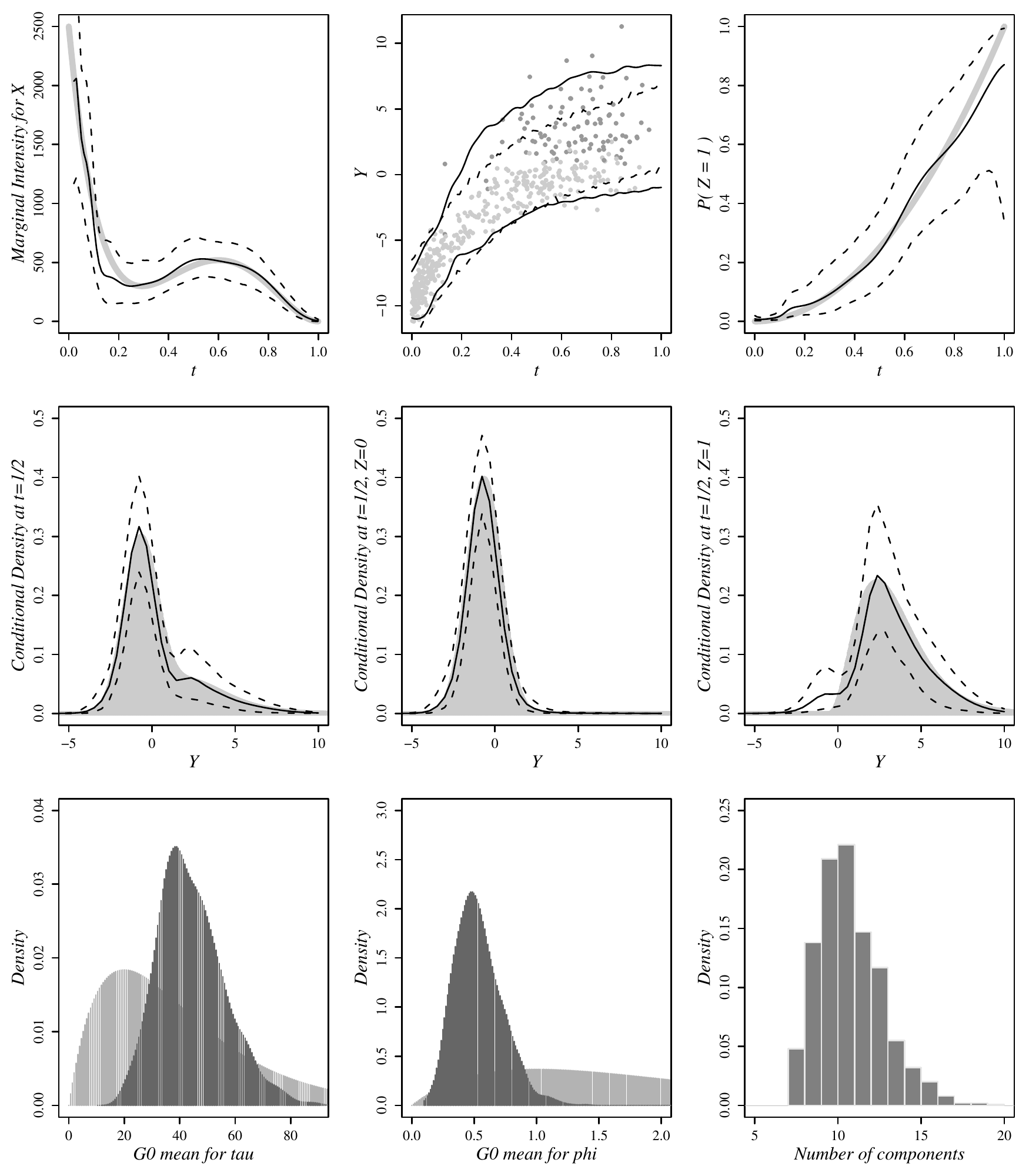}

\caption{Simulation study results.  On top, from left
  to right, we have posterior mean and 90\% interval for the marginal
  intensity $\lambda(t; G)$ (with the true intensity denoted by the grey 
  line), the data (dark grey for $z=1$), and posterior 90\% predictive 
  intervals based on both $h(y \mid t;G)$ (solid
  lines) and GP regression (dotted lines), and posterior mean and 90\%
  intervals for $\mr{Pr}(z=1 \mid t; G)$ (with the true function denoted 
  by the grey line). The middle row has mean and
  90\% intervals for conditional densities for $y$ at $t=1/2$, 
  marginalized over $z$ (left panel) and conditional on $z$ (middle and 
  right panels), with true densities plotted in grey.
  Lastly, the bottom row shows posterior samples for $\beta_{\tau}$ 
  and $\beta_{\phi}$ (dark grey, with priors in the background) and 
  for the number of latent mixture components.}\label{simstudy} 
\end{figure}

Results are shown in Figure \ref{simstudy}.  In the top row, we see
that our methods are able to capture the marginal point intensity and
general conditional behavior for $y$ and $z$; note that the uncertainty 
bounds are based on a full assessment of posterior uncertainty that
is made possible through use of the truncated $G_L$ approximations to 
random mixing measure $G$ (as developed in Section \ref{inference}).
We also fit a Gaussian process (GP) regression model to the $(t,y)$ 
data pairs (using the \texttt{tgp} package for {\sf R} under default 
parametrization) and, in contrast to our approach based on draws from 
$h(y \mid t;G_L)$ as in (\ref{approxcnd}), the top middle panel shows 
the GP model's global variance as unable to adapt to a wider skewed error 
distribution 
for larger $t$ values.

The middle row of Figure \ref{simstudy} illustrates behavior for a
slice of the conditional mark density for $y$, at $t=1/2$, both
marginally and given $z=0$ or $1$.  The marginal (left-most) plot
shows that our model is able to reproduce the skewed response
distribution, while the other two plots capture conditional response
behavior given each value for $z$.  As one would expect,
posterior uncertainty around the conditional mark density estimates is 
highest at the transition from normal to gamma errors. Finally, posterior 
inference for model characteristics is illustrated in the bottom row of Figure
\ref{simstudy}. Peaked posteriors for $\beta_{\tau}$ and $\beta_{\phi}$ show 
that it is possible to learn about hyperparameters of the DP base distribution
for both $t$ and $y$ kernel parameters, despite the flexibility of a DP
mixture. Moreover, based on the posterior distribution for $m$, we note that 
the near to 500 observations have been shrunk to (on average) 12 distinct 
mixture components.

\subsection{Temporal Poisson process with count marks}
\label{coal-mining}

Our second example involves a standard data set from the literature,
the ``coal-mining disasters'' data
\citep[e.g.,][p. 53-56]{AndrewsHerzberg1985}.
The point pattern is defined by the times (in days) of 191 explosions
of fire-damp or coal-dust in mines leading to accidents, involving 10
or more men killed, over a total time period of 40,550 days, from 15
March 1851 to 22 March 1962. The data marks $y$ are the number of
deaths associated with each accident.

This example will compare two different mixture models for marginal location
intensity: a ``direct'' model with beta-Poisson kernels, and a
``transformed'' model with data mapped to $\ds{R}^2$ and fit via
multivariate normal kernels.  The first scheme models data directly
on its original scale, but requires Metropolis-Hastings augmented MCMC
for the beta kernel parameters, 
and dependence between $t$ and $y$ is induced only through $G$.  
The second model affords the convenience of the collapsed Gibbs sampler
and correlated kernels, but on a transformed scale.

Following our general modeling approach, both models use the reference
prior for $\Lambda_{\mc{R}}$ and assume NHPP density form $f(t,y;G)=$ 
$\int \mr{k}(t,y;\theta)dG(\theta)$ with $G \sim
\mr{DP}(\alpha, G_0)$ and $\pi(\alpha)= \mr{ga}(2, 1)$.  The
distinction between the two models is thus limited to choice of
kernel and base distribution. For the direct model,
\begin{eqnarray}\label{truncpois}
  \mr{k}(t,y;\mu,\tau,\phi) &=& b(t;\mu,\tau) \mr{Po}_{\geq 10}(y;\phi),\\
g_{0}(\mu,\tau,\phi) &=& \ds{1}_{\mu \in (0,1)}
  \mr{ga}(\tau^{-1}; 2, \beta_{\tau})\mr{ga}(\phi ; 1, 1/60), \notag
\end{eqnarray}
where $\mr{Po}_{\geq 10}(y;\phi)$ is a Poisson density truncated at
$y=10$, and with $\pi(\beta_{\tau}) = \mr{ga}(1, 1/63)$.  This leads to
prior expectations $\ds{E}[\phi] = 60$ and $\ds{E}[\tau]=$
$\ds{E}[\beta_{\tau}] = 63$ for mean location kernel precision
$(1+\tau)/(\mu(1-\mu))$ $\approx$ $4(1+63)$, which translates to a
standard deviation of $1/16$.  For the transformed model, we take
$\tilde{y}=$ $y-9.5$ and
\begin{eqnarray}
  \mr{k}(t,y;\bs{\mu},\bs{\Sigma}) &=& 
\frac{\mr{N}\left([\mr{logit}(t), \log(\tilde{y})]';
\bs{\mu},\bs{\Sigma}\right)}{\tilde{y}t(1-t)}\\
g_{0}(\bs{\mu},\bs{\Sigma}) 
&=& \mr{N}(\bs{\mu}; (0,2.5)',
10\bs{\Sigma})W(\bs{\Sigma}^{-1};3,\bs{\Omega}), \notag
\end{eqnarray}
with $\pi(\bs{\Omega})=$ $\mr{W}(3, \mr{diag}[10,20])$ for
$\ds{E}(\Sigma)=$ $2/3 \ds{E}(\Omega) = \mr{diag}[1/5,1/10]$
($\mr{logit}(t)$ and $\log(\tilde{y})$ range in (-5,5) and (-1,6),
respectively). Both models were found to be  robust to
changes in this parametrization (e.g., $\ds{E}[\phi] \in
[10,100]$ and diagonal elements of $\ds{E}[\Sigma]$ in $[0.1,1]$).

\begin{figure}[t]
\includegraphics[width=6.25in]{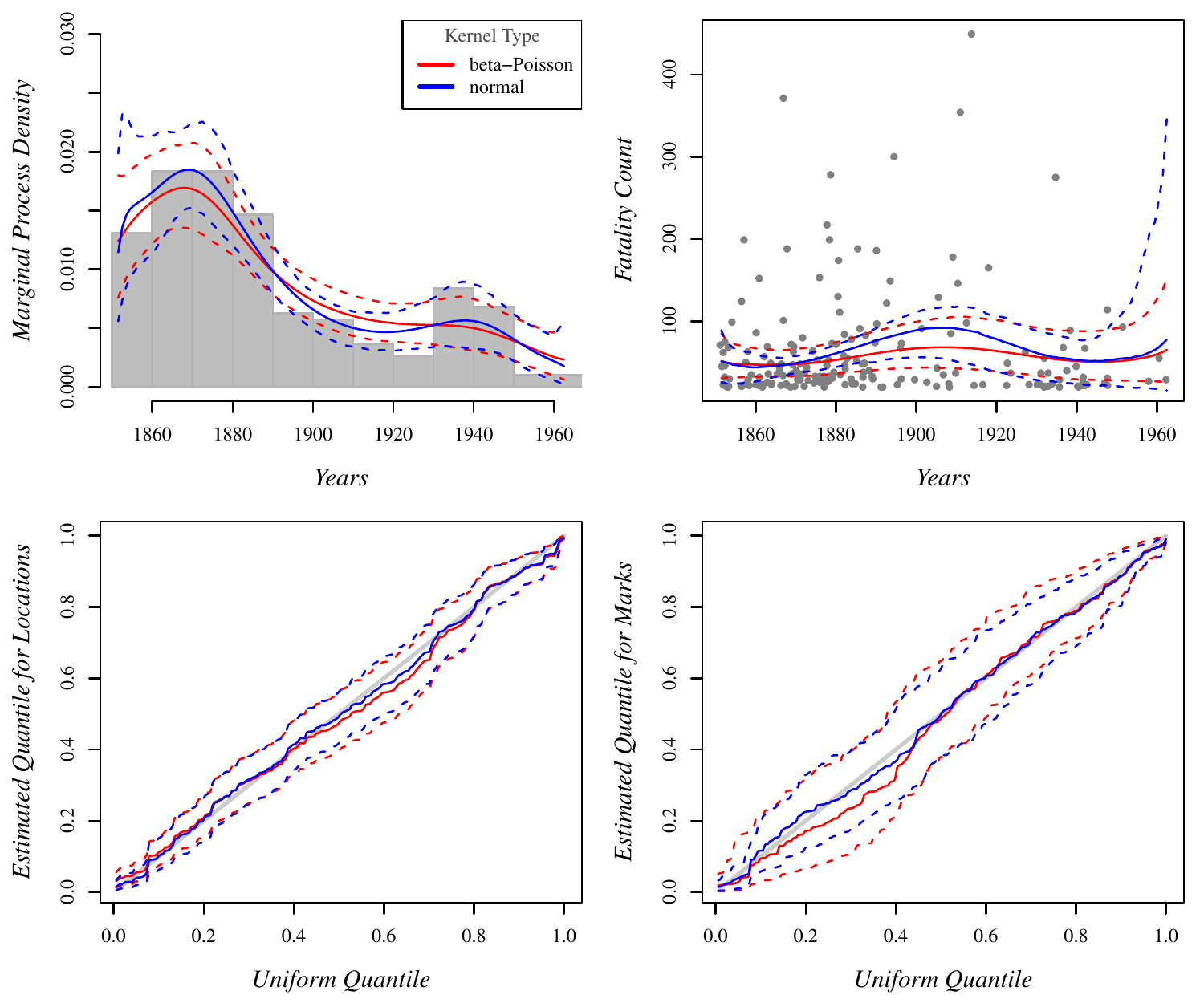} 
\caption{
Coal-mining disasters. Mean and 90\% intervals for (clockwise from top-left): 
marginal density $f(t;G_{L})$ (with data histogram); conditional expected 
count $\ds{E}(y \mid t;G_{L})$ (data counts in grey); and posterior Q-Q plots 
for $\mr{Pr}(y < y_i \mid t_i; G_{L})$ and $\Lambda(t_i;G_{L})$,
respectively.} \label{coal} 
\end{figure}

%
%
Results under both models are shown in Figure \ref{coal}. In the top 
left panel, we see that marginal process density estimates
derived from each model are generally similar, with the
normal model perhaps more sensitive to data peaks and troughs.  There
is no noticeable edge effect for either model. The Q-Q plot in the 
bottom left panel shows roughly similar fit with the normal model 
performing slightly better.  
The top and bottom right panels report inference for the count mark
conditional mean and distribution Q-Q plot. For the beta-Poisson model, 
posterior realizations for $\ds{E}(y \mid t;G_L)$ are obtained using
(\ref{cond-mean}).
The conditional mean calculation for the normal model must account for 
the correlated kernels (and the transformation to $\tilde{y}$), such
that $\ds{E}(y \mid t;G_L) $ is
$$
\frac{9}{2} + 
\left(q_0\sum_{l=1}^L p_l \mr{N}(t; \mu_{lt}, \sigma^2_{lt})
\ds{E}[y \mid t;\vartheta_l] + \sum_{j=1}^mq_j\mr{N}(t; \mu^\star_{jt}, \sigma^{\star 2}_{jt})
\ds{E}[y \mid t;\theta^\star_j]\right)/{f(t;G_L)}
$$
where $\ds{E}[y \mid t,\theta]=$ $\exp\left[\mu_{y} + \rho
  \sigma_{t}^{-2}(t-\mu_{t}) + 0.5(\sigma^{2}_{y} - \rho^{2}
  \sigma_{t}^{-2}) \right]$ with $\bs{\mu}=$ $(\mu_{t},\mu_{y})$ and
$\bs{\Sigma}$ partitioned into variances
$(\sigma^2_{t},\sigma^{2}_{y})$ and correlation $\rho$. Similarly,
uniform quantiles for the conditional mark distribution under the
beta-Poisson model are available as weighted sums of Poisson
distribution functions, while the normal model calculation for
$\mr{Pr}(y < y_i \mid t_i; G_L)$ is as above for $\ds{E}(y \mid
t;G_L)$, but with $\ds{E}[y \mid t,\theta]$ replaced by $\mr{Pr}(y <
y_i \mid t_i; \theta)=$ \linebreak $\Phi\left(\left[\tilde{y}_i -
    \mu_{y} + \rho \sigma_{t}^{-2} (t_i-\mu_{t})\right]
  (\sigma^{2}_{y} - \rho^{2} \sigma_{t}^{-2})^{-1/2} \right)$.  The
estimated conditional mean functions are qualitatively different, with
the poisson model missing the peak at WW1.  Indeed, the corresponding
QQ plot shows that the normal model provides a better fit to this
data; we hypothesize that this is due to the equality of mean and
variance assumed in Poisson kernels, and may be fixed by using
instead, say, truncated negative binomials.

\subsection{Spatial Poisson process with continuous marks}
\label{pine-trees}

Our final example considers the locations and diameters of 584
Longleaf pine trees in a $200\times200$ meter patch of forest in
Thomas County, GA.  The trees were surveyed in 1979 and the measured
mark is diameter at breast height (1.5 $m$), or $dbh$, recorded only
for trees with greater than 2 {\it cm dbh}.  The data,
available as part of the \texttt{spatstat} package for \texttt{R},
were analyzed by \cite{RathCres1994} as part of a space-time survival
point process.  Poisson processes are generally viewed as an
inadequate model for forest patterns, due to the dependent birth
process by which trees occur. However, the NHPP should be flexible 
enough to account for variability in tree counts at a single time point 
and, in this example, we will concentrate primarily on inference for 
the conditional $dbh$ mark distribution.

\begin{figure}[t]
\includegraphics[width=2.25in]{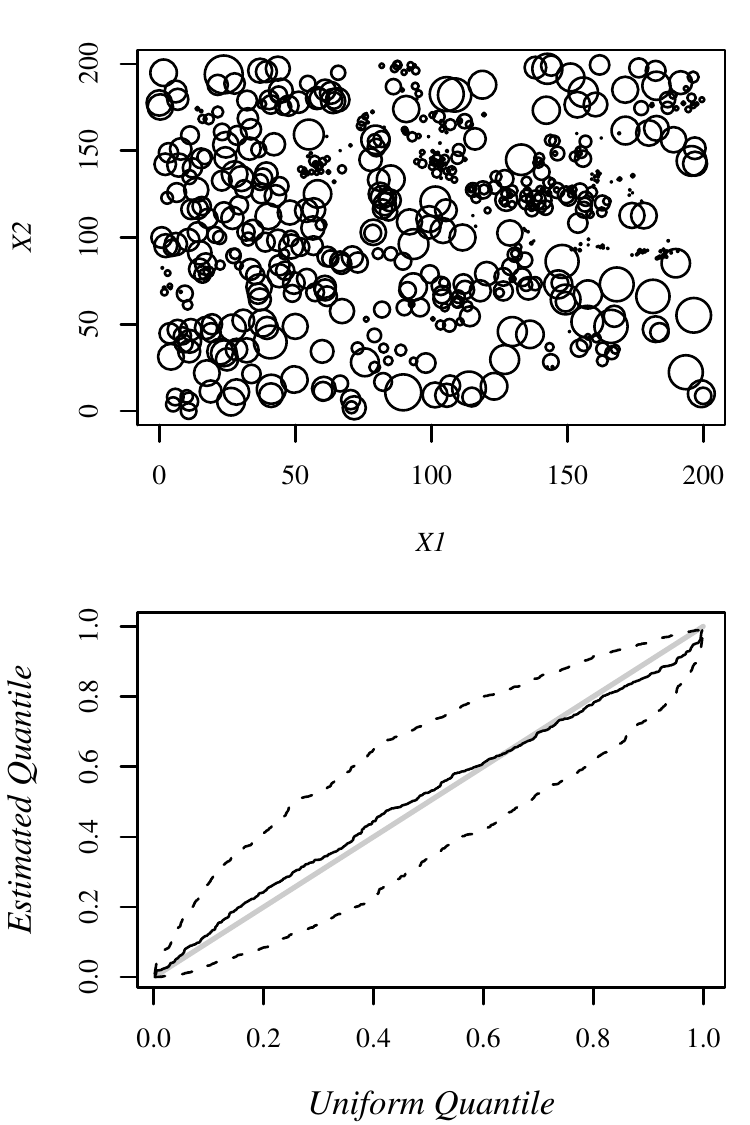}
\includegraphics[width=4.1in]{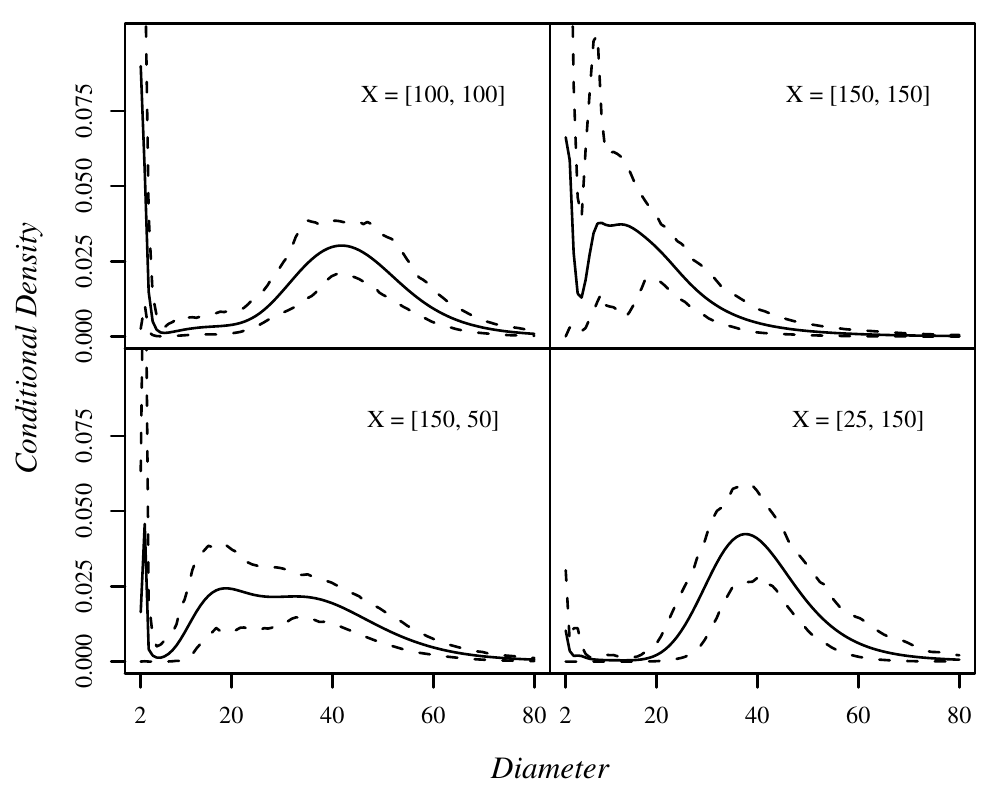}
\caption{Longleaf pines. The left panel has data (point size 
proportional to tree diameter) and a Q-Q plot (mean and 90\% interval) for 
$\int^{y} h(s \mid \bm{x};G_{L})ds$ evaluated at data. The right panel plots 
posterior mean and 90\% intervals for $h(y \mid \bm{x};G_{L})$ at four 
specific $\bm{x}$ values.}\label{pine} 
\end{figure}

To analyse this data set, we  employ a spatial version of the model in
(\ref{mvnmarks}), with tree marks log-transformed to lie on the real
line. Thus, our three-dimensional normal kernel model is
\[
  \phi(\bm{x},y;G) = \Lambda_{\mc{R}} \int \frac{\mr{N}\left(
[\mr{logit}(\bm{x}),\log(y-2)]'; \bs{\mu},\bs{\Sigma} 
\right) } {(y-2) \prod_{i=1}^2 x_i (1-x_i )} d
  G(\bs{\mu},\bs{\Sigma}), 
  \hspace{0.5cm} G \sim \mr{DP}(\alpha,G_{0}).
\]
The base distribution is taken to be $g_{0}(\bs{\mu},\bs{\Sigma})=$
$\mr{N}(\bs{\mu}; (0,0,1)',100\bs{\Sigma})\mr{W}(\bs{\Sigma}^{-1};
4,\bs{\Omega})$, with $\pi(\bs{\Omega})=$ $\mr{W}(4,\mr{diag}[0.1,0.1,0.1,0.1])$.
A $\mr{ga}(2,1)$ prior is placed on $\alpha$. Posterior sampling follows 
the fully collapsed Gibbs algorithm of Appendix A.2.

In this data set, high density clusters of juveniles trees ({\it dbh} 
$<5${\it cm}) combine with the more even dispersal of larger trees to 
form conditional mark densities with non-standard shapes and non-homogeneous
variability. This behavior is clearly exhibited in the posterior estimates
of the conditional density for $dbh$, shown on the right side of
Figure \ref{pine}, at four different locations in the observations
window. Although conditional densities vary in shape over the
different locations, each appears to show the mixture of a diffuse
component for mature trees combined with a sharp increase in density
at low $dbh$ values, corresponding to collections of juvenile trees
(only some of whom make it to maturity).  It is notable that we are
able to infer this structure nonparametrically, in contrast to
existing approaches where the effect of a tree-age threshold is
assumed {\it a priori} \citep[as in][]{RathCres1994}. Finally, the
conditional mark distribution Q-Q plot on the bottom right panel of 
Figure \ref{pine} (based on calculations similar to those in Section 
\ref{coal-mining}) shows a generally decent mean-fit with wide uncertainty 
bands corresponding to the 95\% and 5\% density percentile Q-Q plots.

\section{Discussion} 
\label{summary}

We have presented a nonparametric Bayesian modeling framework for
marked non-homogeneous Poisson processes. The key feature of the
approach is that it develops the modeling from the Poisson process
density. We have considered various forms of Dirichlet process mixture
models for this density which, when extended to the joint
mark-location process, result in highly flexible nonparametric
inference for the location intensity as well as for the conditional
mark distribution. The approach enables modeling and inference for
multivariate mark distributions comprising both categorical and
continuous marks, and is especially appealing with regard to the
relative simplicity with which it can accommodate spatially correlated
marks. We have discussed methods for prior specification, posterior
simulation and inference, and model checking. Finally, three data
examples were used to illustrate the proposed methodology.

The Poisson assumption for marked point processes is what enables us
to separate modeling for the process density from the integrated
intensity. This simplification is particularly useful for applications
involving several related intensity functions and mark
distributions, and is less restrictive than it may at first appear.
For instance, \cite{Taddy2010} presents an estimation of weekly
violent crime intensity surfaces, using autoregressive modeling for marked
spatial NHPPs, and \cite{KBMPO2011} compares  neuronal
firing intensities recorded under multiple experimental conditions,
using hierarchically dependent modeling for temporal NHPPs.

Among the possible ways to relax the restrictions of the Poisson
assumption, while retaining the appealing structure of the NHPP
likelihood, we note the class of multiplicative intensity models
studied, for instance, in \citet{IshwJame2004}. These models for
marked point processes are under the NHPP setting and, indeed, follow
the simpler strategy of separate modeling for the process intensity
and mark density as in the semiparametric framework of Section
3.1. More generally, one could envision relaxing the Poisson
assumption for the number of marks through a joint intensity function
such that the location intensity is not the marginal of the joint
intensity over marks.  Such extensions would however sacrifice the
main feature of our proposed framework -- flexible modeling for
multivariate mark distributions under a practical posterior simulation
inference scheme.  As a more basic extension, our factorization in (1)
could be combined with alternative specifications for integrated
intensity; for example, hierarchical models may be useful to connect
intensity across observation windows.

\section*{Acknowledgements}

The authors wish to thank an Associate Editor and a referee for
helpful comments.
The work of the first author was supported in part by the IBM Corporation 
Faculty Research Fund at the University of Chicago. The work of the second 
author was supported in part by the National Science Foundation under awards 
DEB 0727543 and SES 1024484.

\appendix

\section*{Appendix: Implementation Details for Dirichlet Process Mixture Models}

\subsection*{A.1 ~~Prior Specification}

Prior specification for the DP precision parameter is facilitated by
the role of $\alpha$ in controlling the number, $m \leq N$, of
distinct mixture components.  For instance, for moderately large $N$,
$\ds{E}[m \mid \alpha] \approx$ $\alpha \log \left( (\alpha +
  N)/\alpha \right)$. Furthermore, it is common to assume a gamma
prior for $\alpha$, such that $\pi(\alpha)=$
$\mr{ga}(\alpha;a_{\alpha}, b_{\alpha})$, and use prior intuition
about $m$ combined with $\ds{E}[m \mid \alpha]$ to guide the choice of
$a_\alpha$ and $b_\alpha$.

Specification of the base distribution parameters will clearly depend
on kernel choice and application details, and DP mixture models are
typically robust to reasonable changes in this specification.  First,
the base distribution for kernel {\it location} (usually the mean, but
possibly median) can be specified through a prior guess for the data
center; for example, this value can be used to fix the mean parameter
$\bs{\delta}$ in (\ref{bivarnormal}) or the mean of a normal
hyperprior for $\bs{\delta}$.  In choosing dispersion parameters, note
that the DP prior will place most mass on a small 
number of mixture components, with the remaining components assigned
very little weight and, hence, very few observations.  At the same time, this behavior can
be overcome in the posterior and it is important to not restrict the
mixture to overly-dispersed kernels. Thus, the expectation of the
kernel variance (or scale, or shape) parameters should be specified
with a small number of mixture components in mind, but with low
precision.  For example, again in the context of (\ref{bivarnormal}),
the square-root of the hyperprior expectation for diagonal elements of
$\bs{\Omega}$ can be set at $1/8$ to $1/16$ of a prior guess at data
range, and the precision $\nu$ will be as small as is practical
(usually the dimension of the kernel plus 2).  The factor $\kappa$ is
then chosen to scale the mixture to expected dispersion in $\bs{\mu}$.

Moreover, except when specific prior information about co-dependence is
available, it is best to center $G_0$ on kernel parametrization that
implies independence between variables, such that the mixture is
centered on a model with dependence induced nonparametrically by $G$.
For example, in the model of (\ref{bivarnormal}), we assume zeros in
the off-diagonal elements for the prior expectation of $\bs{\Omega}$,
and this is combined with a small $\nu$ to allow for within-kernel
dependence where appropriate.  A prior expectation of independence
also fits with our general approach of building kernels for mixed-type
data as the product of multiple independent densities.

Note that we have chosen to introduce prior information into the base
measure based on the intuition arising from a small number of large
mixture components and $\alpha$ near zero.  Recent work in
\cite{BushLeeMacE2010} provides a rigorous treatment of
non-informative prior specification, and they advocate a hierarchical
scheme for $\alpha | G_0$ that maintains desirable properties at all
scales of precision.  As the main work here -- use of mixtures for
modeling joint location-mark Poisson process densities -- is
independent of prior and base measure choice, these innovations, as
well as application-specific prior schemes, could potentially be
integrated into our framework.

\subsection*{A.2 ~~Posterior Simulation}

Using results from \citet{Anto1974}, the posterior distribution for
the DP mixture model is partitioned as
$\mr{p}(G,\bs{\theta}^{\star},\bm{s},\alpha,\psi \mid \text{data})=$
$\mr{p}(G \mid \bs{\theta}^{\star},\bm{s},\alpha,\psi)
\mr{p}(\bs{\theta}^{\star},\bm{s},\alpha,\psi \mid \text{data})$,
where $G$, given $\bs{\theta}^{\star},\bm{s},\alpha,\psi$, is
distributed as a DP with precision parameter $\alpha + N$ and base
distribution given by (\ref{polya}).  Hence, full posterior inference
involves sampling for the finite dimensional portion of the parameter
vector, which is next supplemented with draws from the conditional
posterior distribution for $G$ (obtained as discussed in Section
\ref{inference}). A generic Gibbs sampler for posterior simulation
from $\mr{p}(\bs{\theta}^{\star},\bm{s},\alpha,\psi \mid
\text{data})$, derived by combining MCMC methods from
\citet{MacEachern1994} and \citet{EscoWest1995}, proceeds iteratively
as follows:
\begin{itemize}
\item For $i=1,\ldots,N$, denote by $\bm{s}^{(-i)}$ the allocation
  vector with component $s_{i}$ removed, and by $N_{s}^{(-i)}$ the
  number of elements of $\bm{s}^{(-i)}$ that are equal to $s$. Then,
  if $s=s_{r}$ for some $r \neq i$, the $i$-th allocation variable is
  updated according to
$$
\text{Pr}(s_{i} = s \mid \bm{s}^{(-i)},\alpha,\psi,\text{data}) \propto
\frac{N_{s}^{(-i)}}{N - 1 + \alpha} 
\int \mr{k}(\bm{z}_i;\theta^{\star}) p(\theta^{\star} \mid \bm{s}^{(-i)},\psi,\text{data})
d\theta^{\star},
$$
where $p(\theta^{\star} \mid \bm{s}^{(-i)},\psi,\text{data})$ is the
density proportional to $g_{0}(\theta^{\star};\psi) \prod\nolimits_{
  \{ r \neq i: s_{r}=s \} } \mr{k}(\bm{z}_{r};\theta^{\star})$.
Moreover, the probability of generating a new component, that is,
$\text{Pr}(s_{i} \neq s_{r} \, \text{for all} \, r \neq i \mid
\bm{s}^{(-i)},\alpha,\psi,\text{data})$, is proportional to $\alpha (N
- 1 + \alpha)^{-1} \int \mr{k}(\bm{z}_{i};\theta^{\star})
g_{0}(\theta^{\star};\psi) d\theta^{\star}$.
\item 
  For $j=1,...,m$, draw $\theta^\star_j$ from 
  $p(\theta^{\star}_{j} \mid \bm{s},\psi,\text{data}) \propto$
  $g_{0}(\theta^{\star}_{j};\psi) \prod\nolimits_{\{ i: s_{i}=j \}} 
  \mr{k}(\bm{z}_{i};\theta^{\star}_{j})$.
\item 
  Draw the base distribution hyperparameters from 
  $\pi(\psi) \prod_{j=1}^{m} g_{0}(\theta^{\star}_{j};\psi)$, where $\pi(\psi)$
  is the prior for $\psi$. Finally, if $\alpha$ is assigned a gamma 
  hyperprior, it can be updated conditional on only $m$ and $N$ using the 
  auxiliary variable method from \citet{EscoWest1995}.
\end{itemize}
The integrals that are needed to update the components of $\bm{s}$ can
be evaluated analytically for models where $G_0$ is conjugate for
$\mr{k}(\cdot;\theta)$. It is for this reason that conditionally
conjugate mixture models can lead to substantially more efficient
posterior sampling, especially when $\theta$ is high-dimensional. When
this is not true (as for, e.g., beta kernel models or the truncated
Poisson of Equation \ref{truncpois}), the draw for
$\bm{s}$ requires use of the auxiliary parameters,
$\bs{\theta}^{\star}$, sampled as in the second step of our algorithm,
in conjunction with a joint Metropolis-Hastings draw for each
$\theta_i$ and $s_i$ given $\bs{\theta}^{(-i)}$ and
$\bm{s}^{(-i)}$. In particular, we can make use of algorithms from
\citet{Neal2000} for non-conjugate models.


\end{document}